\documentclass[preprint]{aastex62}
\begin{document}

\title{The Rise of ROME. I. A Multiwavelength Analysis of the Star-Planet Interaction in the HD 189733 System}

\author[0000-0001-6987-6527]{Matthew Route}
\affiliation{Department of Astronomy and Astrophysics, the Pennsylvania State University, 525 Davey Laboratory, University Park, PA 16802}
\affiliation{Center for Exoplanets and Habitable Worlds, the Pennsylvania State University, 525 Davey Laboratory, University Park, PA 16802}
\affiliation{Research Computing, Purdue University, 155 S. Grant St., West Lafayette, IN 47907}

\received{May 21, 2018}
\accepted{January 4, 2019}
\published{February 10, 2019}
\submitjournal{The Astrophysical Journal}
\reportnum{ApJ 2019, 872, 79}

\correspondingauthor{Matthew Route}
\email{mroute@purdue.edu}


\begin{abstract}
Several ``hot Jupiter'' host stars appear to exhibit enhanced chromospheric activity, coronal flaring, and starspot development synchronized to their planetary orbits.  These effects have been attributed to tidal and/or magnetic interactions between the star and exoplanet.  The best studied among such systems is HD 189733, which has previously been observed from radio to X-ray wavelengths.  Here, I present $\sim$4.75 GHz Arecibo radio telescope observations of HD 189733 during a fraction of the exoplanet orbital phase range previously associated with enhanced coronal X-ray flaring.  No radio flares from the lower corona were detected. I also leverage \emph{Microvariability and Oscillations of Stars}, Automated Photoelectric Telescope, and Wise photometry to measure the occurences of minima associated with enhanced starspot activity. The phasing of these minima with the exoplanet orbit does not reveal any synchronized active region behavior.  Moreover, for the first time, I combine X-ray, ultraviolet, photometric, \ion{Ca}{2} H\&K, H$\alpha$, and radio observations of HD 189733 to conduct an extensive, holistic reexamination of stellar activity in this system.  Through the presentation of new data, and by merging physical and statistical analytic approaches, I demonstrate that the previously asserted enhancements in activity, allegedly synchronized to certain exoplanet orbital phases, are likely the result of inadequately sampled intrinsic stellar activity from an active star, rather than star-planet interactions.
\end{abstract}
 
\section{Introduction}
During the ``ELODIE metallicity-biased search for transiting Hot Jupiters,'' an exoplanet was detected around the active star HD 189733A \citep{bou05}.  The star's close proximity ($d$=19.25 pc) and brightness ($V$=7.67) have greatly aided study of the system.  Spectroscopic analysis of the 0.82 $M_{\odot}$ host star demonstrated that it has $R_{\ast}$=0.76 $R_{\odot}$, $T_{eff}=5050 \pm 50$K, and [Fe/H]=-0.03 $\pm$ 0.04, resulting in a reclassification of its spectral type as K1-K2.  Astrometry indicates that HD 189733A has a $\sim$M3.5 dwarf companion that orbits in the plane of the sky nearly perpendicular to the exoplanet orbit, with a projected semi-major axis of 216 AU, yielding a $\sim$3200-yr orbit \citep{bak06a}. 

The star is very active, as indicated by its variability classification and chromospheric activity index.  HD 189733A is classified as a BY Draconis star, a group that contains K and M stars that exhibit rotationally-modulated luminosities due to starspots and chromospheric activity.  Many stars in this class have fast rotation periods ($\sim$5 days) since approximately 70\% are members of tidally-locked, short-period binary systems, while the remainder are young stars \citep{sch00}.  HD 189733A has a chromospheric activity index ``$S$-value'' of 0.525, that makes it among the 10\% most active K dwarfs in the California and Carnegie Planet Search Project chromospheric \ion{Ca}{2} emission catalog \citep{wri04,mou07,boi09}.  The $S$-value activity standard was developed during the Mount Wilson dwarf star monitoring campaign.  It measured the quotient of fluxes from 1.09\AA~full width at half maximum (FWHM) triangular bandpasses centered on the \ion{Ca}{2} H\&K lines, and two 20\AA-wide continuum channels that were located blueward of the \ion{Ca}{0} features at 3901.1\AA and redward at 4001.1\AA.

The 1.13$\pm$0.03 $M_{J}$ exoplanet orbits the host star at a distance of 0.031 AU (8.77 $R_{\ast}$) in a 2.2185733$\pm$0.0000019 day orbit \citep{win07,boi09}. $BVRI$ photometry enabled a precise determination of the planetary radius, $R_{P}=1.154\pm0.033 R_{J}$ \citep{bak06b}.  Among the more significant findings from the exoplanet was the construction of a thermal map that showed a $\sim$250 K temperature contrast between day and nightsides, as measured by the \emph{Spitzer} Infrared Array Camera (IRAC) \citep{knut07}.  The planet has an evaporating exosphere of hydrogen gas that overfills its Roche lobe \citep{lec10}.  \emph{Hubble Space Telescope (HST)} transmission spectroscopy from infrared to ultraviolet (UV) wavelengths have revealed the exoplanet's atmospheric composition and indicated the presence of a dusty atmospheric haze \citep{pon13}.

These system properties have spurred research and vigorous debate into the potential existence of a star-planet interaction between HD 189733A and its exoplanet.  \citet{cun00} hypothesized that tidal and magnetic interactions between host stars and exoplanets with semi-major axes $a\lesssim 0.1$ AU may result in enhanced magnetic activity, such as increased chromospheric, transition region, and coronal emission.  Tidal effects may lead to increased dynamo activity in stars with convective zones, and the interaction of the planetary magnetosphere with stellar active region magnetic fields may result in enhanced coronal magnetic reconnection.  \citet{shk08} monitored chromospheric \ion{Ca}{2} H\&K emission from the HD 189733 system and reported the detection of enhanced flaring ostensibly caused by a star-planet interaction.  Following up on this potential discovery, \citet{pil10,pil11,pil14,pil15} conducted X-ray and far ultraviolet (FUV) observations of HD 189733A and reported flaring synchronized to the orbital period of the exoplanet at every observing epoch.  This led \citet{pil14,pil15} to propose that HD 189733A actively accretes from its ``hot Jupiter'' companion, and hosts a persistent active region that migrates across the stellar surface in phase with the orbital motion of the exoplanet.  These studies, together with additional observations described below, make the HD 189733 system the exemplar of exoplanetary systems alleged to have star-planet interactions.

In the following sections, I examine observations of the HD 189733A/b system that span radio to X-ray wavelengths and attempt to mold them into a comprehensive narrative of the system's stellar activity.  Throughout this work, I employ a holistic, multiwavelength approach to the study of activity in this exoplanet system, and also compare the photospheric, chromospheric, and coronal activity of HD 189733A to what is known about activity on other stars, especially the Sun.  Section 2 provides a brief introduction to observational signatures of stellar activity, and describes the motivations behind my analysis at radio and optical wavelengths.  Section 3 describes my radio observations of HD 189733A that searched for evidence of strong flares driven by magnetic reconnection in the system, as well as my measurement of published photometric time series to determine the occurence and phases of active regions.  Section 4 presents the results of these investigations, and descibes their position in the larger context of stellar activity.  In Section 5, I describe and re-evaluate \ion{Ca}{2} H\&K, X-ray, UV, {\bf radio}, photometric, and H$\alpha$ observations of the HD 189733 system, with subsections that focus on each wavelength regime in turn.  These subsections are presented roughly in chronological order, so that the reader may follow the evolution of the study of stellar activity in this system.  In Section 6, I analyze the statistics of this entire ensemble of multiwavelength observations, and conclude in Section 7.  These aggregated results demonstrate that while \citet{shk08} first asserted that an inhomogenous and variable stellar magnetic field interacting with the exoplanet magnetic field quasi-periodically induces enhanced activity, the ensemble of multiwavelength observations do not support this hypothesis.

\section{A Primer on Stellar Activity}
To aid in the evaluation of these claims, it is helpful to briefly review the many observational hallmarks of stellar activity.  Stellar activity is driven by the localized emergence of magnetic flux from the solar interior, and the interaction of these magnetic fields.  It includes the following phenomenon (described in \citet{pri14} unless otherwise noted).
\begin{itemize}
	\item \emph{Granular magnetic loops}, are the smallest scale magnetic flux emergence structures found in the photosphere, with fluxes $\gtrsim$10$^{16}$ Mx and short lifetimes.
	\item \emph{Ephemeral regions} have random orientations and are distributed uniformly across the solar photosphere, with magnetic fluxes fluxes $\sim$10$^{19}$ Mx.  They are manifest by X-ray bright points in the corona.
	\item \emph{Starspots} on the Sun are photometric dark spots 250 to 2,000 K cooler than the surrounding photosphere, with magnetic fields $B$=1.5-6 kG.
	\item \emph{Faculae} are bright regions in the photosphere, and are associated with enhanced chromospheric emission in H$\alpha$ and \ion{Ca}{2} H \& K.  Clusters of faculae are called plages \citep{sch00}.
	\item \emph{Active regions} represent $\sim$10$^{23}$ Mx concentrations of magnetic flux, with mean fields of $B\sim$100 G.  They are manifest by dark sunspot groups in the photosphere and bright plages of chromospheric emission.  They are generally found in two activity bands, located north and south of the equator.  Many flares occur within active regions.
	\item \emph{Flares} result in increased emission throughout the electromagnetic spectrum including $\gamma$-rays, X-rays, UV, $UBVR$ photometry, H$\alpha$, H$\delta$, H$\gamma$, \ion{Ca}{2}, \ion{Mg}{2}, infrared, and radio \citep{haw91,ben02}.
	\item \emph{Jets} are associated with magnetic reconnection, including ubiquitous \ion{Ca}{2} jets in the chromosphere, and coronal X-ray/extreme UV (EUV) jets \citep{lang09}. 
	\item \emph{Prominences} are located in the corona, and appear as H$\alpha$-emitting loops as viewed from the solar limb, but dark H$\alpha$-absorbing filaments as viewed against the disk.
	\item \emph{Coronal Mass Ejections} (CMEs) rapidly expel plasma from the stellar atmosphere, with associated radio, white light, EUV, and soft X-ray emission.
	\item \emph{Radio bursts} are classified as Types I, II, III, IV, centimeter and milliecond. They span frequencies of $\sim$10 kHz to $\sim$100 GHz with durations of $\sim$ms to days.  They are created by various flare and shock processes in the corona \citep{lang09}.
\end{itemize}

\subsection{Context and Motivation for ROME}
The detection of X-ray flares from HD 189733A that are generated either through magnetic reconnection events intrinsic to the primary star, or that may be induced by a star-planet interaction, leads to the possibility that radio emission associated with these events may be detected.  Studies of solar and stellar magnetic reconnection provide the model for this supposition.  The twisting of magnetic field lines causes reconnection events, which accelerate electrons both toward and away from the stellar surface.  These coronal electrons spiral along magnetic field lines and emit bremsstrahlung-dominated hard X-rays and radio waves via the mildly relativistic gyrosynchrotron, plasma, and electron cyclotron maser emission (ECM) mechanisms.  The electrons that impact the chromosphere lower in the atmosphere deposit their kinetic energy into the denser plasma, thereby producing soft X-rays and enhancing chromospheric emission \citep{ben10,whi11}.

Radio observations provide valuable information that complement observations of flares and additional types of stellar activity at other wavelengths.  Radio emissions come from the rarefied layers of the stellar atmosphere, and therefore probe flare conditions in the coronae of stars that would otherwise be inaccessible to remote observation.  These observations allow us to directly examine the physics of nonthermal electron acceleration regions \citep{ben17}.  Measurement of the flux density and circular polarization fraction at radio wavelengths can be used to determine the emission mechanism(s) involved, wave propagation conditions, and reveal the magnetic field strength of the emitting region (e.g., \citet{pic08}).  The sign of the Stokes V signature determines the direction of the magnetic field, which can be used to trace large scale magnetic topology (e.g., \citet{ben10}), or, in spatially-resolved images, reveal changes in magnetic field direction within active regions (e.g., \citet{bog09}).  Radio emission mechanisms constrain the electron density in the emitting region (e.g., \citet{rou12}).  Finally, the drifting of radio emissions in dynamic spectra can reveal plasma motions within a magnetic field (e.g., \citet{ost08}).

Previous radio observations of the HD 189733 system sought low-frequency, ECM-induced radio emission from the auroral regions of the exoplanet, similar to that found at Jupiter \citep{tre06,lec09,smi09,lec11}.  Instead, I searched for radio emission that may be associated with stellar flares or active regions and may be induced by star-planet interactions, as part of the ROME (Radio Observations of Magnetized Exoplanets) program.  The system was observed at higher frequencies than previously used because $\sim$GHz frequencies are typical for the detection of radio bursts emitted by gyrosynchrotron, and potentially, ECM mechanisms associated with stellar flares and CMEs in the lower corona \citep{tre06}.

\subsection{Photometric Reanalysis}
In addition to the search for flaring radio emission, I also analyzed published photometric time series to examine how active regions on HD 189733A may be correlated with other signatures of potential star-planet interactions.  Active regions are small areas of the stellar surface with strong, concentrated magnetic fields that host most of the stellar activity.  At visible wavelengths, they are manifest as starspot groups, which darken the stellar photosphere due to their cooler temperatures.  Typically, regions consist of two starspot groups: in the leading group starspots are geographically concentrated and of a single magnetic polarity, while the trailing group is more spread out and of the opposite polarity.  Thus, the magnetic fields within an active region are generally closed.  The emergence of magnetic flux in these regions is subject to magnetic reconnection that result in plasma heating, flares, and CMEs.  Plasma heating in the chromosphere results in upflows of plasma in magnetically confined coronal loops that produce X-ray, extreme UV (EUV), soft X-ray, and radio emission \citep{asc05}.

Since \citet{pil14,pil15} asserted that an active region migrates across the stellar surface in phase with the exoplanet orbit, we can test this hypothesis by searching for persistent starspot-hosting active regions that reappear at particular orbital phases.  The starspots in these active regions would appear as local minima in the photometric light curve.  Similarly, since active regions are sites of enhanced chromospheric activity, including flaring, the location of photometric minima will also allow us to test the hypothesis, suggested by \citet{shk08}, of increased chromospheric flaring induced by star-planet interactions at a preferred orbital phase.  This investigation will, to my knowledge, constitute the first photometric search for signatures associated with star-planet interactions.

\section{Observations}
\subsection{Radio Observations}
I observed the HD 189733 system using the 305 m William E. Gordon radio telescope at Arecibo Observatory on 2011 September 7, for $\sim$7.2 ks, the time it takes for the target to transit the fixed dish. I monitored the system with the C-band receiver operating at 4.25-5.25 GHz with the same instrumental setup that has been used in my surveys of magnetized brown dwarf radio emission \citep{rouphd,rou13,rw16}.  The antenna gain and system temperatures were approximately 8 K Jy$^{-1}$ and 30 K, respectively, at low zenith angles\footnote{``C-Band,'' available at http://www.naic.edu/$\sim$astro/RXstatus/Cband/Cband.shtml}.  The C-band receiver has a half-power beamwidth of $\sim$1\arcmin in both azimuth and zenith angles.  The receiver collects dual-linearly polarized signals that the Mock spectrometer\footnote{``The `Mock Spectrometer,'' http://www.naic.edu/\%7Eastro/guide/node9.html} processes into full Stokes parameters.  The Mock spectrometer consists of seven individually tunable field-programmable gate array (FPGA) equipped fast Fourier transform (FFT) boxes, each with a 172 MHz bandpass divided into 8192 channels that yield a frequency resolution of $\sim$20 kHz.  The observing run consisted of 12 science scans collected at 0.1 s temporal resolution.  Each 10-min science scan is bracketed by 20 s of calibration-on/calibration-off scans that leverage a local oscillator.   

A natively developed IDL software pipeline, as described in \citet{rouphd}, processes the signals from the Mock spectrometer.  These routines compute the \emph{IQUV} Stokes parameters, perform flux calibration and bandpass corrections, iteratively reduce radio frequency interference (RFI), and resample the data to $\sim$80 kHz spectral and 0.9 s temporal resolutions.  Data analysis consists of examining Stokes V dynamic spectra (time-frequency spectrograms) for bursts of left- or right- circularly polarized emission that do not correspond to previously identified RFI patterns.  Examples of these spectra can be found in \citet{rou13,rw16}.  Due to the broadband nature of gyrosynchotron and ECM emission from stars and brown dwarfs, bursts of radio emission from HD 189733A should appear in more than one box, which aids in the dismissal of false positives.

I transformed the Modified Julian Dates (MJDs) of my observations into Heliocentric Julian Dates (HJDs), then into host star rotational ($E_{Rot}$) and exoplanet orbital ($E_{Rot}$) cycles using the ephemerides provided in \citet{far17}:
\begin{equation} T_{0} = HJD~2453629.389 + 12~E_{Rot},\end{equation}
\begin{equation} T_{0} = HJD~2453629.389 + 2.218575~E_{Orb},\end{equation}
The observations spanned HJD 2455811.508894 to 2455811.597663 and probed $\phi_{orbit}$=0.568-0.608 ($\phi_{orbit}$=0.0 during exoplanet transit), which includes a fraction of the orbital phase range for which \citet{pil10,pil11,pil14,pil15} potentially observed enhanced X-ray flaring.  The experimental setup is only sensitive to rapidly varying Stokes V circular polarization fractions of $\gtrsim$10\% due to confusion limitations and calibration uncertainty in Stokes I total intensity.  However, this circular polarization sensitivity should be adequate to detect emission from gyrosynchrotron, ECM, and plasma emission mechanisms (e.g., \citet{ste01,top10}).

\subsection{Photometric Temporal Analysis}
I analyzed the time series from three photometric data sets, including 21 days of \emph{Microvariability and Oscillations of Stars} (\emph{MOST}) satellite photometry collected in 2006 August \citep{mill08}, $\sim$31 days of \emph{MOST} data from 2007 July to 2007 August \citep{boi09}, and Automated Photoelectric Telescope (APT) and Wise Observatory photometric data from 2009 October to 2009 December and 2010 May to 2010 June \citep{sing11}.  \emph{MOST} photometry was obtained in Direct Imaging mode through a single broadband filter (3500-7000\AA), that was binned in intervals of 21s \citep{mill08} or 101.43 mins \citep{boi09}.  APT, located at Fairborn Observatory, measured Str\"{o}mgren $b$ and $y$ photometry on approximately nightly timescales.  $R$ filter photometry acquired via the Wise Observatory were measured approximately every second night.  From these three data sets, I measured the onset and termination of the local minima in the light curves, which were averaged to produce the time at which the active region was in the center of the disk.  I then used Equations 1 and 2 to compute the rotational and orbital cycles associated with these active regions, as listed in Table 1.  The inherent precision of the data sets makes it unnecessary to convert the temporal measurements to HJD for comparison with other data sets.
	
\section{Results}
\subsection{Radio Results}
No radio bursts were detected during my observations, which have a 3$\sigma$ detection sensitivity of 1.158 mJy in Stokes V.  Although the theoretical 1$\sigma$ sensitivity is $\sim$0.15 mJy over a $\sim$1 GHz bandpass, the erratic presence of RFI across individual boxes reduces this to a 1$\sigma$ sensitivity of 0.386 mJy in the cleanest box.  This limit on sensitivity corresponds to the ability to observe flares with energies $\nu~L_{\nu}\geq 2.183\times 10^{24}$ erg s$^{-1}$ for istotropic (nonbeamed) emission at 4.25 GHz.  These results are 5-50$\times$ more sensitive than previous searches for flaring radio emission, with order-of-magnitude smaller integration times.

\subsubsection{Radio Observations throughout the Main Sequence and Beyond}

The Sun is by far the best-studied star at radio wavelengths and has been observed since the invention of radar (for a review, see \citet{pic08} and references therein).  Its proximity enables the detection and characterization of weak radio activity at high temporal resolution, and spatially resolved images enable the correlation of radio activity to that observed at other wavelengths.  The maximum radio emission detected during Solar Cycle 23, which peaked in 2000, was from a CME associated with the eruption of a small, M2.3 flare from active region AR 8649 on 1999 July 28.  The peak recorded radio emission was 42,000 s.f.u. (solar flux units; 1 s.f.u.=$10^{-19}$ erg s$^{-1}$ cm$^{-2}$ Hz$^{-1}$) at 606 MHz, which corresponds to $\nu~L_{\nu}=7.158\times 10^{21}$ erg s$^{-1}$ \citep{che11}.  The sensitivity threshold derived for HD 189733A is therefore significantly higher than the most powerful radio emissions from the Sun measured in the literature.

The Sun, however, is a much less active star than HD 189733A, which indicates that comparison with other stars may better gauge its stellar activity.  In recent years, stellar radio variability has been discovered stretching from the B2V HR 7355 (HD 182180) \citep{let17}, through the M8 DENIS 1048-3956 at the bottom of the main sequence \citep{bur05}, and beyond to as late as the T6.5 brown dwarf 2MASS J10475385+2124234 \citep{rou12}.  Observations of stellar activity at radio wavelengths, nevertheless, remain decidedly rare and are not well characterized.  For example, flare flux density distributions generally have been measured to have power-law index $\alpha\sim$2 at various wavelengths (e.g., \citet{haw14}).  Although a similar index has been determined for solar radio emission (e.g., \citet{cro93}) and is hypothesized to be applicable to ultracool dwarfs (spectral types $\geq$M7) and brown dwarfs \citep{rou17}, the precise value of this index and its applicability to other stellar and substellar objects at radio wavelengths have yet to be demonstrated.

Another potential means of comparison is to attempt to leverage flare statistics accumulated from other stellar surveys at other wavelengths.  Simultaneous multiwavelength observations of stellar flares have revealed that the association of radio flares and the distribution of their energy relative to their $\gamma$-ray, X-ray, UV, and optical flare counterparts are complicated.  Although large flares are generally associated with Type III radio bursts on the Sun, nearly 1/3 of \emph{Reuven Ramaty High Energy Solar Spectroscopic Image (RHESSI)} hard X-ray flares are unassociated with radio bursts.  On the other hand, a number of radio flares occur in the absence of an X-ray flare \citep{ben17}.  In addition, there are few examples of solar flares observed simultaneously throughout the electromagnetic spectrum (e.g., \citet{can80,ben02}), and multiwavelength campaigns of other stars have not yielded straightforward results.  For example, \citet{ost05} observed the dM4.5e star EV Lacertae simultaneously in the radio with the Very Large Array (VLA), with optical photometry and spectroscopy provided by McDonald Observatory, with \emph{HST}/Space Telescope Imaging Spectrograph (STIS) UV spectroscopy, and with X-ray coverage provided by \emph{Chandra X-Ray Observatory} High Energy Transmission Grating Spectrometer (\emph{Chandra}/HETGS) and Advanced CCD Imaging Spectrometer (ACIS).  For the most part, they found that radio enhancements were uncorrelated with optical, UV, and X-ray enhancements, although certain flares were correlated in a subset of these wavelength regimes, or in rare cases, all of them.  At the stellar/substellar boundary, \citet{ber08} simultaneously observed the M9 TVLM 513-46546 at X-ray (\emph{Chandra}/ACIS), UV (\emph{Swift} Ultraviolet/Optical Telescope, UVOT), H$\alpha$ (Gemini), and radio (VLA) wavelengths and also found that while very few flaring events were correlated in radio and X-rays, the overwhelming majority of flares were uncorrelated across wavelengths, and even \emph{anti}correlated in H$\alpha$.  These considerations make the problem of estimating radio luminosities and occurrence rates for flares on HD 189733A, and the comparison of my radio results with diagnostics of stellar flares at other wavelengths, such as amplitude, energy, and duration diagrams and flare frequency diagrams (FFDs) (e.g., \citet{sil16}) nontrivial.

\subsubsection{Incoherent Emission Processes: The G\"{u}del-Benz Relationship}
However, the radiation mechanisms potentially involved in any nonthermal emission processes provide some guidance in the interpretation of these results.  Nonthermal emission, such as hard X-rays and radio, can be caused by either coherent or incoherent processes.  We will start with an estimate of the anticipated radio luminosity from incoherent processes.  The G\"{u}del-Benz relationship describes the correlation between nonthermal, incoherent, gyrosynchrotron radio emission and the thermalized X-ray emission in flares witnessed on F-M stars, including the Sun.  The G\"{u}del-Benz relationship is given by
\begin{equation} {L_{X} \over L_{R}}=\kappa \times 10^{15.5\pm 0.5}~[Hz],\end{equation} 
where $L_{X}$ is the X-ray luminosity, $L_{R}$ is the radio luminosity, and $\kappa$ varies by type of star but, in this case, is of order unity \citep{gud93,ben94}.  Using the peak X-ray flare luminosity \citet{pil14} reported ($L_{X}=1.950\times 10^{28}$ erg s$^{-1}$), the G\"{u}del-Benz relationship yields a maximum peak radio flare luminosity at 4.25 GHz of $8.3\times 10^{22}$ erg s$^{-1}$.  However, it is important to note that the G\"{u}del-Benz relationship expected $L_{X}$ to measure soft X-rays down to 0.1 keV, whereas the \emph{XMM-Newton} bandpass that \citet{pil14} used spanned only 0.3-8 keV.  This difference in expected energy ranges could result in an underestimate of $L_{X}$ for HD 1897333A by nearly two orders of magnitude \citep{ben94}.
	
A reference star more similar to HD 189733A may better estimate the X-ray and radio luminosities involved in incoherent processes.  $\epsilon$ Eri is another active K2V star with similar mass (0.86 $M_{\sun}$), radius (0.74 $R_{\sun}$), and rotation period (11.68 days) to HD 189733A, if somewhat smaller mean magnetic field strength (10-20 G), and should therefore have similar levels of activity \citep{jef14}.  \citet{nes02} give the maximum X-ray luminosity for $\epsilon$ Eri as $L_{X}=2.09\times 10^{29}$ erg s$^{-1}$ as measured by the \emph{Chandra} Low Energy Transmission Grating Spectrometer (\emph{Chandra}/LETGS) over a 0.07-2.5 keV bandpass.  Inserting this luminosity into the G\"{u}del-Benz relationship, we would anticipate a maximum flare luminosity $\nu~L_{\nu}$=$8.9\times 10^{23}$ erg s$^{-1}$ for radio emission at 4.25 GHz.

Another means to estimate $L_{R}$ is by leveraging the X-ray luminosity range determined for other BY Dra class stars.  Figure 11 of \citet{ben10} gives their peak X-ray luminosity range as $L_{X}\sim10^{27.5}$-$10^{30}$ erg s$^{-1}$ which corresponds to 4.25 GHz radio luminosities of $\nu~L_{\nu}\sim 4\times 10^{21}$-$4\times 10^{24}$ erg s$^{-1}$.  Thus, these considerations indicate that for flares generated by incoherent gyrosynchrotron emission, as is typical in many stellar flares, my observations exclude only the most powerful flares.

\subsubsection{Coherent Emission Processes: The Electron Cyclotron Maser}
Alternatively, coherent emission processes, such as plasma and ECM emission, may preferentially beam radio emission toward the Earth as their active regions rotate into view.  Indeed, solar flares larger than C5 class always radiate some coherent radio emission (\citet{ben17} and references therein).  The observational signatures of ECM emission include high circular polarization fractions, high brightness temperatures, and fine structure in the radio dynamic spectrum.  For example, during Effelsberg observations of the dM3.5e flare star AD Leonis, \citet{ste01} detected an entirely right-circularly polarized, irregular sequence of pulses with peak flux density $F_{\nu}\sim$100 mJy.  Assuming that the emitting region was the size of the stellar disk, they computed a brightness temperature $T_{B}\sim 5\times 10^{10}$ K, while for reasonable flare loop sizes $T_{B}\geq 3\times 10^{13}$ K.  Using this peak flux density and the distance to the star, $d=$4.85 pc, the ECM mechanism would produce a flare with an \emph{isotropic} (nonbeamed) radio luminosity $\nu~L_{\nu}=1.365\times 10^{25}$ erg s$^{-1}$.  Obviously, once the effects of beaming are taken into account, the real flare luminosity is much less.  If such flares occurred on HD 189733A, their radio flux densities would be readily detectable at Arecibo Observatory.  Thus, these observations exclude the presence of extremely large gyrosynchrotron flares and modest-sized coherent ECM flares in the lower corona of HD 189733A at $\phi_{orbit}$=0.568-0.608 (Figure 1).

\subsubsection{Detection Probability of Flaring}
Using the multiwavelength observations of flares discussed in Section 5 and recorded in Table 2, we can compute an average time between flare flux peaks.  This leads to a time scale of $\sim$49600 s between flares.  Observation of the system for $\sim$7.2 ks at radio wavelengths, therefore, yields a detection probability for a single flaring event of $\sim$15\%.  This value does not take into account the duration of the flares, which on K dwarfs have median white-light durations of $\sim$3.5 hours \citep{wal11} or the ability to detect flares that follow an undetermined flux energy distribution.  Alternatively, \citet{wal11} found that the median time spent in a flaring state among 373 K-M flare stars in the \emph{Kepler} Quarter 1 data release was 1-4\%.  Again, although the temporal and energetic properties of stellar flares at other wavelengths are well known, it is nontrivial to convert optical, UV, and X-ray flare luminosities into anticipated radio luminosities, since the partitioning of flare energy into different wavelength bands varies from flare to flare and is known for only a handful of solar flares (e.g., \citet{can80}).

\subsection{Photometric Temporal Analysis Results}
It is clear from my temporal analysis of the photometric data (Table 1) that starspot features persist over several rotation periods and are better correlated with rotation phase than the orbital phase of the exoplanet.  If the star-planet interaction from the exoplanet periodically, or quasi-periodically, enhances stellar activity, it may be supposed that this enhancement would result in an increase in starspots at certain orbital phases that would be manifest as photometric minima synchronized to a particular orbital phase.  However, this effect is not observed, as can readily be discerned in Tables 1 and 3.  We will analyze the statistics of these results both separately, and as part of the ensemble of multiwavelength observations of HD 189733A activity, in Sections 6.1 and 6.3, respectively.

\section{A Comprehensive Discussion of Prior Multiwavelength Results}
In this section, I review the chronology of the alleged discovery of star-planet interactions in the HD 189733 system.  I begin with the \ion{Ca}{2} H\&K observations that first generated interest in the system and recount additional instances of stellar activity found throughout the literature.  For the first time, I investigate whether the multitude of observations from across the electromagnetic spectrum are consistent with enhanced, exoplanet-induced magnetic activity, or whether the observed activity is not atypical for an active star such as HD 189733A.  All activity presented in this section is listed in Table 2.

\subsection{Description and Reassessment of \ion{Ca}{2} H\&K and Zeeman Doppler Imaging Observations}
\citet{shk03} initially investigated the star-planet interaction by observing the chromospheric \ion{Ca}{2} H\&K emission lines (3933.7, 3968.5\AA) of HD 179949.  They found enhanced emission that appeared to coincide with the location of the subplanetary point ($\phi_{orbit}\sim$0).  The team then investigated the star-planet interaction in several additional exoplanet systems, including HD 189733. Keen interest in the magnetic properties of this system stems from the \emph{abstract} of \citet{shk08}, which states \emph{``variability in the transiting system HD 189733 is likely associated with an active region rotating with the star; however, the flaring in excess of the rotational modulation may be associated with its hot Jupiter.''}  The team collected four nights of spectra from the Echelle Spectropolarimetric Device for the Observation of Stars (ESPaDOnS) on the Canada-France-Hawaii Telescope (CFHT).  They computed the mean absolute deviation (MAD=$N^{-1}\Sigma_{i}|data_{i}-mean|$ for $N$ spectra) for residuals of the normalized spectra containing the \ion{Ca}{2} K line.  From this analysis, they found that \emph{``no correlation is seen between the planet's orbit and the residuals to the rotational modulation,''} as depicted in their Figure 11.  However, they mention that the \ion{Ca}{2} K MAD shows a \emph{``clear increase in very short ($\leq$30 minutes) activity at $\phi_{orb}\sim$0.8,''} which is manifested in their Figure 12 as a maximum among their four nightly data sets at $\phi_{orbit}\sim$0.8.

Comparing Figures 11 and 12 reveals the perplexing behavior of this increase in the \ion{Ca}{2} K MAD, that measures chromospheric variability and flaring.  If the maximum in MAD at $\phi_{orbit}\sim$0.8 represents an increase in chromospheric activity, there should be a rapid increase in residual \ion{Ca}{2} K emission above that indicated by the rotationally fitted model.  However, close inspection of their Figure 11 indicates that the night with the most variation (and hence, largest MAD) is the second set from the left, and shows a large $\sim$6\% \emph{decline} in \ion{Ca}{2} K emission below the best-fit model to the rotationally modulated chromospheric emission.  Such a precipitous decrease in chromospheric emission is difficult to explain; for comparison, it is even greater in magnitude than the $\sim$5\% decline in \ion{H}{1} Lyman-$\alpha$ (1215.6\AA) flux absorbed by the evaporating exosphere of the planet as it transited the stellar disk \citep{lec10}.  Stellar flares generally exhibit the opposite behavior and result in a large increase in \ion{Ca}{2} K emission, as demonstrated by simultaneous \emph{Kepler} and Apache Point Observatory spectroscopic observations of flares on the dM4e star GJ 1243 \citep{sil16}.  On the Sun, a weakening in chromospheric \ion{Ca}{2} H\&K emission is found at very quiet regions, while active regions exhibit increasing emission \citep{lin17}.  The peak-to-peak scatter in integrated residual \ion{Ca}{2} K flux, barring this single low measurement, is $\sim$0.03, the same as for the other nights.  \citet{shk08} analyzed several other spectral lines that are typically indicative of chromospheric activity, including \ion{Ca}{2} H, Ca IRT (8662\AA), H$\alpha$ (6562.8\AA), and \ion{He}{1} D3 (5875.6\AA).  Among these, only the other \ion{Ca}{2} lines were correlated with \ion{Ca}{2} K activity; the H$\alpha$ and \ion{He}{1} D3 lines either were poorly correlated, or were difficult to measure, respectively.  Therefore, of the five spectroscopic activity indicators observed, only the three \ion{Ca}{0} measurements supported the assertion of enhanced chromospheric activity.  Rather than the spectroscopic data demonstrating enhanced stellar chromospheric flaring phased with the orbital motion of HD 189733b, the low \ion{Ca}{2} emission at $\phi_{orbit}\sim$0.8 is instead consistent with an absence of flaring and the rotation of a \emph{quiet} region into view (or an active region out of view), in contradiction to the hypothesis presented.

Subsequently, an arsenal of spacecraft and instruments conducted detailed, multiwavelength analyses of the HD 189733 system. \citet{mou07} used ESPaDOnS to collect 10 nights of spectra over two observing epochs, in order to leverage Zeeman signatures to infer the star's large-scale magnetic topology.  Noting that previous work found rotation periods for HD 189733 ranging from 11.7 to 13.4 days, they suggested that differentially rotating bright and dark active regions at various latitudes observed over multiple epochs could explain this behavior (e.g., \cite{win07}).  They determined that \ion{Ca}{2} H\&K core emission flux can vary by $\pm$10\% on timescales of $\sim$1 hr and that \emph{``longer-term variations in core emission are apparently not related to the orbital phase,''} but were correlated to stellar rotation phase.  Although their results do not support the presence of a star-planet interaction, they cautioned that denser temporal sampling of the magnetic activity as a function of both orbital and rotational phases would be required to conclude that the activity is entirely independent of HD 189733b's orbital phase. 

Similarly, \citet{boi09} obtained 55 high-resolution spectra of HD 189733A with the SOPHIE spectrograph at the Observatoire de Haute-Provence from 2007 July to 2007 August and analyzed activity in \ion{Ca}{2} H\&K, H$\alpha$, and \ion{He}{1} lines.  They computed a battery of diagnostics, including the cross-correlation function (CCF) for all spectral lines, CCF bisector velocity spans (V$_{span}$) that measure line asymmetry, chromospheric line spectral indices, ``observed-minus-calculated'' (O-C) radial velocity residuals, and the measurement of \emph{MOST} photometry. Spectral indices for the \ion{Ca}{2}, H$\alpha$, and \ion{He}{1} lines were computed using $Index=F_{SL}/(F_{L}+F_{R})$, where $F_{L}$ ($F_{R}$) denotes the continuum flux in narrow windows on the left- (right-) hand sides of the spectral line for which a flux is measured ($F_{SL}$).  The wavelength windows for the $F_{SL}$, $F_{L}$, and $F_{R}$ components all varied based on the spectral feature being probed.  The noisier \ion{Ca}{2} and \ion{He}{1} indices were much less useful for analysis than the H$\alpha$ index, which was correlated with the stellar rotation but uncorrelated with exoplanet orbital motion.  The O-C residuals, V$_{span}$, \emph{MOST} photometry, and the location of two minima in the \ion{He}{1} index light curve indicated, as a first approximation, the presence of a single active region on the stellar surface that was phased with the stellar rotation period alone.  Thus, \citet{boi09}'s detailed, multi-faceted analysis of HD 189733's chromospheric activity indicators also does not support the star-planet interaction hypothesis.

\citet{far10} collected 44 spectra over 27 nights from 2007 June to 2008 June using the ESPaDOnS and NARVAL spectropolarimeters, the latter of which is located at the Telescope Bernard Lyot.  They found that the H$\alpha$ and \ion{Ca}{2} H\&K indicators were modulated on 11.6-14.4 and 11.9-12.2 day timescales, respectively.  Differential rotation rates ranging from 11.94$\pm$0.16 days at the equator to 16.53$\pm$2.43 days at the poles explained both data sets well.  To search for lower-amplitude signals from a potential star-planet interaction, they subtracted the best-fitting rotation solutions to these activity indicators and independently searched the H$\alpha$ and \ion{Ca}{2} H\&K residuals for periodicities.  They hypothesized that any effect of the star-planet interaction would be observed as a periodicity of 2.5-2.7 days, corresponding to the beat period between the stellar rotation and exoplanet orbital periods.  Only a single H$\alpha$ observing epoch was within 1$\sigma$ of the beat period, while none of the \ion{Ca}{2} H\&K emission residuals conformed to this periodicity.  The residual emission varied considerably at every orbital phase, indicating that intrinsic variability, not a star-planet magnetic interaction, was the cause.

\citet{cze15} witnessed a flare on 2012 July 1 with the Very Large Telescope (VLT-UT2) Ultraviolet and Visual Echelle Spectrograph (UVES) during their study of center-to-limb variations of the HD 189733A disk.  The onset of the flare coincided with a steep rise in \ion{Ca}{2} H\&K flux at midtransit ($\phi_{orbit}=$0.005), followed by a decay phase that lasted until the end of the observation.  This flare helps us to measure the intrinsic variability of HD 189733A, and to evaluate whether its activity is enhanced near the subplanetary point ($\phi_{orbit}=$0.0).

During their Multiwavelength Observations of an eVaporating Exoplanet and its Star (MOVES) program (2013 June to 2015 July), \citet{far17} used Zeeman Doppler Imaging (ZDI) of HD 189733A from the NARVAL and ESPaDOnS spectropolarimeters to measure a mean magnetic field strength that varied from 18 G in 2006 August to 36-42 G in 2015 July.  Most of the energy was allocated to the toroidal field, while the weaker, radial field was found to be largely nonaxisymmetric.  Although they did not witness a magnetic field polarity reversal associated with a magnetic activity cycle, they noted that the field evolved significantly on a monthly timescale.  The nonuniform magnetic topology and temporal evolution of the stellar magnetic field caused order-of-magnitude variations in the magnitude of the field at the distance of the exoplanet, but the field never exceeded $B\sim$0.1 G (e.g., their Figure 6).  By comparison, the magnetic field strength at the top of solar flare loop arcades, derived from radio brightness temperatures and hard X-ray spectral electron energy spectral indices, is $B\sim$200-500 G \citep{whi11}.  Thus, ZDI results indicated that HD 189733A has strong intrinsic variability and provided an estimate of the magnetic field strengths that would be involved in any star-planet interaction.  However, any exoplanet-induced magnetic perturbations are quite tiny compared to the magnetic field strengths required to initiate stellar flares.

In summary, despite five separate observing campaigns that searched for enhanced \ion{Ca}{2} H\&K chromospheric emission that would be indicative of enhanced flaring or an exoplanet-synchronized active region, only a single campaign found evidence to support the hypothesis of a star-planet interaction at a preferred orbital phase.  Three additional ZDI campaigns (\citet{mou07,far10,far17}) determined that the magnetic field of HD 189733A evolves significantly on short timescales, but they found no evidence for a star-planet interaction.

\subsection{Description and Re-evaluation of X-ray Observations}
\citet{pil10} analyzed two epochs of \emph{XMM-Newton} observations of HD 189733 conducted with the European Photon Imaging Cameras (EPIC) pn at 0.3-8.0 keV.  The $\sim$55 ks 2007 April 17 observation was recorded during primary eclipse (transit), while the $\sim$35 ks 2009 May 18-19 observation occurred during secondary eclipse.  Although the authors noted a modest flare accompanied by spectral hardening at $\sim$30 ks ($\phi\sim$0), alterations in the X-ray flux are unremarkable compared to variability throughout the time series.  \citet{pil10} witnessed an $F_{X}=1.3\times 10^{-13}$ erg s$^{-1}$ cm$^{-2}$ X-ray flare at $\phi_{orbit}$=0.54 during the 2009 observation, which they speculated may \emph{``arise from active regions for which the activity is triggered by the complex magnetic interaction that should be present between the star and planet.''}  \citet{pil10} linked their result to the potential \ion{Ca}{2} H\&K detection of a star-planet interaction, noting that \emph{``\citet{shk08} has also suggested some excess of flaring activity on HD 189733 in phase with the planet rotation superimposed to the main long lasting active regions on the stellar surface.''}  Follow-up $\sim$39 ks \emph{XMM-Newton} EPIC observations on 2011 April 30 recorded another X-ray flare beginning at $\phi_{orbit}$=0.52 with similar energy properties to the 2009 flare, but longer duration \citep{pil11}.  They concluded that \emph{``the recurrence of such flares is explained by the following scenario: an active region is present at the same location on the stellar surface of both observations.  The magnetic interaction with the planet is inducing a flaring activity in this region.''}

\emph{Swift} X-Ray Telescope (XRT) observations simultaneous with UV exosphere observations (Section 5.3) detected an X-ray flare in the 0.3-3 keV energy band $\sim$8.5 hours prior to transit, which tells us about the intrinsic variability of the source \citep{lec12}.  \citet{pop13} observed HD 189733A during six transits from 2011 July 5 to 2011 July 18 with \emph{Chandra}/ACIS, and noted little X-ray variability for an active star, and no large flares.  Although they measured transit depths varying from 2.3\% to 9.4\%, in excess of the broadband transit depth of 2.4\%, they argued that this effect was due to the existence of an extended exoplanetary atmosphere, not the transit of a permanent, large starspot or associated coronal hole.  Thus, the observations of \citet{pop13} do not support the existence of enhanced activity near $\phi_{orbit}\sim$0, as might be expected in both tidal and magnetic models of the star-planet interaction.

Additional \emph{XMM-Newton} observations on 2012 May 7 detected a small X-ray flare at $\phi_{orbit}$=0.57, and a larger X-ray flare that resembled those spotted in 2009 and 2011, at $\phi_{orbit}$=0.64 during a $\sim$62 ks exposure \citep{pil14}.  \citet{pil14} analyzed the temporal behavior of all X-ray flares reported in the literature, which represented $\sim$370 ks of observations from \emph{Chandra}, \emph{XMM-Newton}, and \emph{Swift}.  They reported \emph{``both an excess of flares at 0.45-0.65 phase interval and a deficit in the 0.9-0.1 range with significance $>$1$\sigma$, while the \emph{Swift} flare is within 1$\sigma$ of the expected number of flares.''}  I note that the \emph{Swift} flare orbital phase is misreported in their Figure 9 and Table 3 as $\phi_{orbit}$=0.2, whereas it should be $\phi_{orbit}$=0.84 since it occurs \emph{before} transit \citep{lec12}.  This led \citet{pil14} to hypothesize that: \emph{``When the planet is at phase 0.55, a region on the surface leading about 70-75 deg from the sub-planetary point emerges from the limb.  This region could be the site of the X-ray flares and responsible for the enhanced chromospheric activity observed in \ion{Ca}{2} lines.  Given the motion of the planet, it is expected that the active region should travel at the same rate of the orbital velocity of the planet on the stellar surface and produce strong flares while on the visible face of the star, or about when the planet is in the range $\phi$=0.55-0.15.''}  Thus, \citet{pil14} theorize that an active region synchronized to the exoplanet's orbital motion moves across the stellar surface.  However, it is noteworthy that the overabundance in flaring that they found at orbital phases $\phi_{orbit}$=0.52-0.65 does not correspond well to the putative chromospheric activity enhancement \citet{shk08} found at $\phi_{orbit}\sim$0.8.  Furthermore, the alleged flaring excess that \citet{pil14} described is not statistically significant, nor does it fit logically into the scenario that they proposed, since an active region can hardly be responsible for a \emph{decline} in flaring.  Moreover, the case for an overabundance of magnetic activity at certain orbital phases is not compelling in light of the accumulated multiwavelength observations reported in the literature (Table 2, Figure 2).

\subsection{Description and Reassessment of Ultraviolet Observations}
Following the Lyman-$\alpha$ detection of atomic hydrogen escaping from the exoplanet HD 209458b \citep{vid03}, \citet{lec10} searched for similar signatures of an exosphere during the transit of HD 189733b.  Using the \emph{HST} Advanced Camera for Surveys (ACS) Solar Blind Camera, they observed three transits from 2007 June 10 to 2008 April 24.  They measured a 5.05$\pm$0.75\% Lyman-$\alpha$ transit depth and detected a $\sim$30 min flare at midtransit of the third epoch in \ion{C}{2} (1335.1\AA). This flare helps us to quantify the intrinsic variability of HD 189733A.

\citet{pil15} observed HD 189733 with the \emph{HST} Cosmic Origins Spectrograph (COS) on 2013 September 12 ($\phi_{orbit}$=0.500-0.626) in search of FUV signatures of the star-planet interaction that corresponded to their X-ray flares.  They noted two distinct flaring episodes of the lines \ion{C}{2} (1334.5-1335.7\AA), \ion{N}{5} (1238.8-1242.8\AA), \ion{Si}{2} (1264.7\AA), \ion{Si}{3} (1201.0-1206.6\AA), and \ion{Si}{4} (1393.8,1402.8\AA) at $\phi_{orbit}$=0.525 and $\phi_{orbit}$=0.588 with significance of 8-10$\sigma$.  They attributed these features to \emph{``a stream of gas evaporating from the planet is actively and almost steadily accreting onto the stellar surface, impacting at 70$\degr$-90$\degr$ ahead of the subplanetary point.''}  This assertion of a steady accretion stream further elaborates on the persistent active region scenario proposed in \citet{pil14}.  Although the X-ray observations of \citet{pil10,pil11,pil14} are generally consistent with these scenarios, they are excluded by observations of the same orbital phase range at other wavelengths.  A more thorough evaluation of this accretion scenario will be forthcoming in a later publication.

\subsection{Description and Reinterpretation of Radio Observations}
Previous searches for radio emission from the HD 189733 system sought beamed, ECM auroral emission from its orbiting exoplanet.  The system was thought to be particularly promising since its ``hot Jupiter'' companion would encounter a dense stellar wind that would result in more powerful auroral emission at low frequencies.  It was also thought that observing a decline in radio emission from the system during secondary eclipse would yield strong proof that exoplanet aurorae contributed to any radio emission.  \citet{smi09} first observed the HD 189733 system for $\sim$5.7 hrs on 2007 April 21 at 327 MHz in dual polarization mode using the Robert C. Byrd Green Bank Telescope (GBT).  The 3$\sigma$ sensitivity for the entire observation was 81 mJy.  Although they detected 1.5$\sigma$ and 2.0$\sigma$ brightenings in 330-333 MHz and 335-339 MHz subbands, respectively, they noted that these were statistically insignificant.

\citet{lec09} conducted a $\sim$7.7 hr observation of the system on 2008 August 14 at 244 MHz (LL polarization only) and 614 MHz (RR polarization only) using the Giant Metrewave Radio Telescope (GMRT).  They reported a 3$\sigma$ sensitivity threshold of 2 mJy at 244 MHz and 0.16 mJy at 614 MHz for the entire observation.  This sensitivity decreases to $\sim$50 mJy and $\sim$5 mJy, respectively, for a default integration time of 16.78 s, which is more useful for detecting flaring behavior.  \citet{lec11} followed up on this effort with a $\sim$9 hr observation of HD 189733 on 2009 August 15 at a central frequency of 148 MHz.  They collected RR/LL polarimetry that had a 3$\sigma$ sensitivity of $\sim$25 mJy over an integration time of 339 s.  A search for flaring activity on timescales of 1 to 30 mins revealed no detections during either campaign.

These three observing campaigns were all centered on $\phi_{orbit}\sim$0.5 and would have been capable of detecting flares with $\nu~L_{\nu}=1.4\times 10^{24}$ to $1.2\times 10^{25}$ erg s$^{-1}$.  Although the frequencies probed were aimed at determining the cyclotron frequency of auroral emission of the exoplanet, it is interesting to note that they are each useful in the context of solar radio emissions.  Type III bursts from the Sun typically occur at frequencies $\nu\sim$300 MHz, while Type IV bursts have been observed at $\nu\sim$170 MHz and $\nu\sim$600 MHz \citep{pic08}.  However, since every search for flaring radio emission at $\phi_{orbit}$=0.41-0.61 ended in failure, these observations, together with my own, do not lend support to the notion of enhanced stellar activity at the orbital phase range \citet{pil10,pil11,pil14,pil15} indicated.

\subsection{Description and Reanalysis of Photometric Observations}
In a search for starspot-induced, quasi-periodic flux variations that would enable the measurement of the host star's rotation period, \citet{win07} measured Sloan Digital Sky Survey $z$-band, Str\"{o}mgren $b$ and $y$, or $V$ and $I$-band photometry that spanned eight transits and 2 years of out-of-transit observations, as part of the Transit Light Curve project (2005-2006).  The 1.3\% peak-to-peak variation in their photometric light curves indicated a starspot covering fraction of $\sim$1\% that changed substantially over each 13.4$\pm$0.4 day rotation and thus lacked permanent features.

\citet{pon07} monitored HD 189733A with \emph{HST}/ACS for three epochs spanning 2006 May 22 to 2006 July 14.  They detected and characterized two photometric flux increases of $\sim$0.1\% and $\sim$0.04\%, respectively, that occurred during transit at $\phi_{orbit}$=0.002 during epoch 1 and $\phi_{orbit}$=0.007 during epoch 2.  The first increase indicated that the planet occulted a large, cool starspot or starspot group that spanned $\sim$80,000 km in longitude and $\sim$12,000-165,000 km in latitude, depending on the temperature contrast between the spots and the surrounding photosphere.  The second event consisted of a transit of a smaller active region.  Photometry accumulated during the third epoch potentially indicated the transiting of a small starspot group during egress.  These observations, therefore, rule out a persistent active region at the subplanetary point that may be induced by a magnetic or tidal star-planet interaction.  Instead, they demonstrate that the starspots evolve with time and may have a decay timescale similar to the Gnevyshev-Waldmeier rule for sunspots \citep{sol03}.

\citet{mill08} analyzed six transit light curves collected during a 21-day \emph{MOST} broadband optical photometric monitoring campaign of HD 189733A.  They found no indications that the exoplanet transited starspots $>$0.2$R_{J}$.  They also compared their results to a ``\citet{cro07} model,'' which allegedly fit the HD 189733A light curve with two large permanent spots rotating at a 11.73-day period.  Such a model was attractive since it would indicate the presence of star-planet-interaction-induced permanent magnetic structures on the stellar surface.  \citet{mill08} found that the transit events they observed would either occult the northern permanent spot in the missing \citet{cro07} model in 4/6 transits, which was not observed, or would transit the large starspot south of the equator, which could not be discerned due to uncertainties in the observations and unaccounted-for model.  Although \citet{mou07,pon07,mill08,shk08,sing11} reference a two-starspot model for HD 189733A in \citet{cro07}, these citations appear to be in error, as the paper neither mentions the detection of starspots nor presents a starspot model.  Furthermore, \citet{mill08} state that this model assumed a stellar rotational inclination angle $i$=59$\degr$, whereas measurement of the Rossiter-McLaughlin effect yielded $i=$85.5$\degr$ \citep{tri09}.  Thus, mistaken citations, together with the measurement of the stellar rotational inclination angle, rule out the presence of the large, permanent starspots \citet{mill08} depicted that may have been related to a star-planet interaction.

\citet{sing11} leveraged \emph{HST}/STIS near-UV-optical spectra to characterize starspots near transits ($\phi_{orbit}$=0.995) that occurred on 2009 November 20 and 2010 May 18.  The spectra indicated spot temperatures of $\sim$4250 K, which would constrain the size of the large active region in \citet{pon07} to be $\sim$12,000 km in latitude.  These spot characteristics enabled them to reproduce the photometric light curve with 6-96 spots that cumulatively decrease the stellar flux by 1-2.8\%.  Higher spot numbers may be supported by the exceeding rarity of spot-free \emph{HST} observations.  These observations demonstrated that starspots observed at the subplanetary point are of normal size and evolve with time.  Furthermore, they suggest that HD 189733A likely hosts many starspots and is intrinsically active.

\citet{pon13}, through a comprehensive analysis of the HD 189733b transmission spectra stretching from UV to infrared wavelengths, confirmed a 1-2\% starspot fraction limit based on the failure of starspots to leave spectral signatures on stellar \ion{Na}{0} and \ion{Mg}{0} H lines.  Their analysis also constrained any large, permanent starpots to be located at the poles and to have anomalous spectra that differ from a cooler stellar photosphere.  The temporal properties of the minima in photometric light curves (Section 4.2), which correspond to dark starspots, and the analysis of \citet{pon13} do not support the hypothesis that a persistent active region of high magnetic field strength exists near $\phi_{orbit}$=0.52-0.65 or $\phi_{orbit}\sim$0.8 \citep{shk08,pil14,pil15}, nor do they support the permanent, large two-spot model of \citet{mill08}.

\subsection{Description of H$\alpha$ Observations}
In addition to the H$\alpha$ observations of the HD 189733 system by \citet{shk03,boi09,far10} described above, \citet{cau17a} conducted a series of short-cadence, high-resolution, out-of-transit H$\alpha$ observations from 2013 July 4 to 2016 September 19 to examine stellar variability on short timescales.  These observations covered $\phi_{orbit}\sim$0.0-0.25, 0.45-0.60, and 0.90-1.05.  They found that relative changes in the H$\alpha$ core flux varied most from median values before and after exoplanet transit.  They hypothesized that this effect could be due to star-planet interactions near the subplanetary point, or could indicate the presence of absorbing circumplanetary material near the exoplanet, in agreement with the evaporating exosphere model of \citet{lec10}.

\citet{cau17b} examined seven H$\alpha$ transits collected by the HARPS spectrograph at La Silla and the HIRES spectrograph at Keck Observatory from 2006 August 21 to 2015 August 4.  Only two data sets showed significant correlation between H$\alpha$ and \ion{Ca}{2} H\&K emission, as occurs for solar activity \citep{liv07}.  Thus, absorption by a planetary exosphere is more likely to contribute to the observed variability than stellar activity.  However, a stellar activity model that includes spots, filaments, and faculae concentrated at latitudes $\sim\pm40\degr$ that provide a covering fraction of $\sim$5-10\%, could also explain the observed H$\alpha$ variations. Such a model resembled the chromospheric activity of the active Sun and would not suggest any enhancement from star-planet interactions.  Thus, although these H$\alpha$ observations do not permit us to evaluate the hypothesis of enhanced chromospheric activity at $\phi_{orbit}\sim$0.8 \citep{shk08}, they do exclude persistent faculae at $\phi_{orbit}\sim$0 \citep{cun00} and $\phi_{orbit}\sim$0.52-0.65 \citep{cun00,pil11,pil14,pil15}.

\subsection{Multiwavelength Phenomenology Summary}
Independently, observations of stellar activity on HD 189733A do little to constrain the various claims of enhanced photospheric, chromospheric, transition region, and coronal activity.  However, a holistic, multiwavelength approach to star-planet interactions strongly constrains the issue. 
 
\begin{enumerate}
\item Although \citet{cun00} hypothesized that ``hot Jupiters'' as a class may induce enhanced magnetic activity near the subplanetary point ($\phi_{orbit}\sim$0), the X-ray and photometric results rule out a persistent active region near this orbital phase.  H$\alpha$ emission may indicate the presence of stellar activity near the subplanetary point, but this activity is similar to that found on the active Sun and would not be anomalous for an even more active star such as HD 189733A.  Furthermore, a more likely explanation for H$\alpha$ variability near $\phi_{orbit}\sim$0 is the presence of the evaporating exosphere of the nearby exoplanet.
\item Although \citet{shk08} suggested that HD 189733A displays enhanced \ion{Ca}{2} H\&K emission at $\phi_{orbit}\sim$0.8, the only support for this claim is to be found in the \emph{abstract} of that paper and not within the data itself.  Furthermore, five independent \ion{Ca}{2} H\&K observing campaigns, three ZDI campaigns, and photometric monitoring campaigns failed to find any anomalous flaring or spotting at this orbital phase.
\item Although \citet{pil10,pil11,pil14,pil15} reported that HD 189733A exhibits enhanced X-ray and FUV activity at $\phi_{orbit}\sim$0.52-0.65, H$\alpha$ observations do not indicate any abnormal chromospheric activity, radio observations did not detect any significant coronal flaring, and photometric monitoring failed to find persistent or recurring active regions at these preferred orbital phases.
\item Although the results of \citet{mill08} may be suggestive of persistent spots associated with a star-planet interaction, other photometric results exclude their model.  Furthermore, the inclination angle, $i$, that the model relies on has been falsified by measurements of the Rossiter-McLaughlin effect.
\end{enumerate}

Moreover, the various claims of enhanced stellar activity in individual wavelength regimes at preferred orbital phases are excluded by multiwavelength observations that fail to observe related signatures of enhanced activity (i.e. flaring, starspots, faculae). Furthermore, the competing claims are not consistent with each other.

\section{Statistical Analyses of Multiwavelength Observations}
Setting aside for the moment the temporal, multiwavelength, and physical inconsistencies of the alleged enhanced activity described previously, we can evaluate these claims through statistical analysis.  For the first time, I investigate whether the hypothesis of star-planet interactions in the HD 189733 system is supported by flare \emph{and} starspot activity statistics.  I first describe the general setup of the statistical analyses and then apply them to the new photometric data presented in Table 1, to the activity recorded in the literature but excluding the new radio and photometric data presented here, and, finally, to the entire ensemble of observed stellar activity in this system.

Solar activity is largely a stochastic process, although flares, photospheric starspots, and chromospheric plages are clustered in active regions. Flares in an active region may also trigger additional nearby, ``sympathetic'' flaring as described by ``avalanche theory'' (e.g., \citet{lu91}).  However, the active regions themselves are often randomly distributed in longitude across the solar surface, and some magnetic activity occurs outside these regions \citep{ben10}.  The relationship between starpots, active regions, and flaring has also been studied for a number of stars, thanks in no small part to the wealth of photometric data available from \emph{Kepler}.  The majority of these studies found that the phase difference between the photometric minima, caused by a visible stellar disk with more active regions and starspots than average, and the maximum amplitude of flares is uniformly distributed, particularly for the most powerful flares \citep{hun12,roe13,haw14,sil16,doy18,roe18}.  I therefore model the flaring and spotting activity as a Poisson process that occurs randomly in orbital phase, but at some definite average rate.  No effort is made to simulate other parameters in the system, such as time between flares, their energy distribution, and variation in instrument sensitivity, due to the uncertainties entailed in these processes and described in detail in Section 4.1.  We can then compare the number of measured instances of stellar activity in each orbital phase range with the number of events that would be counted in each bin if the activity is randomly distributed across the surface in the absence of a star-planet interaction.

In Tables 3, 4, and 5, I divide the planet's orbital phases into five equally divided segments, each spanning 0.2 in orbital phase, and choose phase ranges such that all activity near transit ($\phi_{orbit}\sim$0.0) would fall into the same bin, all activity near the X-ray flares reported by \citet{pil10,pil11,pil14,pil15} ($\phi_{orbit}\sim$0.52-0.65) would be binned together, and any activity related to the claimed enhanced \ion{Ca}{2} H\&K activity of \citet{shk08} ($\phi_{orbit}\sim$0.80) would be binned together.  Note that the latter two alleged episodes of enhanced activity compose $\sim$1/3 of an orbit and are not binned together since this would reduce our analysis to orbital phase ranges that represent transit phases, unobserved phases, and anti-subplanetary point phases. Since I assume a Poisson distribution as a function of orbital phase, the expected number of flares and starspots ($\mu$) can be computed from the fraction of observing time spent in each orbital phase bin times the total number of events recorded, with standard deviation $\sigma=\sqrt\mu$ \citep{tay97}.  

\subsection{Analysis of Active Region Photometry}
Using photometric data from \emph{MOST}, APT, and Wise, I found the timings, rotational cycles, and orbital cycles of 17 photometric minima that correspond to pronounced active regions in the photosphere.  Leveraging the method described above on the data in Table 1 results in the statistical analysis listed in Table 3.  Significantly, this analysis shows that the measured number of active regions does not differ from active regions randomly distributed in orbital phase by more than 0.87$\sigma$.  When the data set is reduced so that photometric minima are located at least one rotational period apart, there is still no statistically significant preferred orbital phase range.  Thus, the star-planet interaction does not enhance starspot formation in the photosphere of HD 189733A.

However, the interpretation of photometric data is not without its difficulties.  Decreases in photospheric flux represent a higher-than-average covering fraction of cool, dark starspots and active regions.  On the other hand, gradual increases in the photometrically measured photospheric flux denote regions with relatively few starspots and active regions.  Sudden sharp, $\sim$tens-of-minutes-long increases in photometric flux denote white-light stellar flares (e.g., \citet{doy18}).  However, a stellar photosphere entirely devoid of starspots and active regions is indistinguishable from an axisymmetric distribution of these features from photometry alone.  A stellar photosphere may be dotted with a multitude of small starspots and active regions, but because of their spatial distribution, it will be unclear to photometric instruments whether flares occurred near starspots.  Yet a darker hemisphere should, overall, have more spots and active regions, thus increasing the likelihood of flaring activity.  Thus, HD 189733A may well have flaring active regions located at preferred orbital phases, but sometimes particularly dense starspot distributions confound our observations. 

Another consideration is that starspots evolve over both short and long timescales.  Large sunspots evolve in accordance with the Gnevyshev-Walkmeier rule and endure for up to several months; starspots on HD 189733A may evolve similarly and should therefore exist for several rotation periods (e.g., \citet{sol03}).  New active regions are also likely created during this time interval, while others disappear.  Differential rotation will alter the photometry of a particular rotational phase as starspot groups are carried to different longitudes at differing rates as a function of their latitudes (e.g., \citet{hal76}).  Finally, long-term ($\gtrsim$year), phased photometric measurements of active regions are likely not comparable due to the existence of stellar magnetic activity cycles that extend from F stars with convective outer envelopes potentially to fully convective stars \citep{bal95,rou16}.  These alter the degree of spottedness and the latitudes at which they emerge.  All of these considerations serve to complicate the analysis presented here.  However, if either temporal evolution of the starspots or their spatial distribution can effectively confound this analysis, it implies that the star-planet interaction is exceedingly weak.

The association of stellar flares with starspots and active regions is the subject of debate.  On the Sun, the best-studied laboratory for the observation of stellar activity, \citet{lee14} noted that the probability of observing flares markedly increased with sunspot area, which would imply that the darkest hemisphere of a star would likely emit the most flares.  On the other hand, studies of correlations between starspots and flares on other stars have not produced straightforward results.  \citet{wal11}, in their analysis of flares from 373 K and M dwarfs in the \emph{Kepler} Quarter 1 data, found that larger photometric variability is correlated with greater energy release from flares.  \citet{sil16} examined flare occurrence rates and energies as a function of rotation phase in \emph{Kepler} photometry of the dM4e star GJ 1243 and found no clear correlation.  However, they suggested that the star may host a long-lived polar starspot that, given the low inclination angle $i\sim$32$\degr$, would result in little change of the spot distribution with rotational phase.  \citet{doy18} analyzed \emph{Kepler} K2 photometry of 34 M0-L1 dwarfs and found that both low- and high-energy flares were uncorrelated with photometric minima and maxima.  It is important to note that in each of these studies \emph{Kepler} photometry is only sensitive to high-energy flares that would correspond to the most powerful observed on the Sun.  Nevertheless, despite these concerns about relations between starspots and flares, \citet{pil14,pil15} explicitly posited an active region synchronized to a particular orbital phase, and the chromospheric variability of \citet{shk08} may be related to active regions.  Yet this statistical analysis demonstrates that the photometric minima do not correspond to either of these hypothesized orbital phase ranges.

\subsection{Analysis of Flare and Starspot Activity Reported in the Literature}
The orbital phases of the X-ray, UV, \ion{Ca}{2} H\&K, H$\alpha$, photometric, and radio activity listed in Table 2 can likewise be analyzed for the signature of a star-planet interaction.  For the purposes of this analysis, the claimed activity of \citet{shk08} and \citet{pil10} is included (see Table 2 notes), but the H$\alpha$ variability detected by \citet{cau17a} is assumed to be from the evaporating exosphere and is therefore excluded, as are my radio and photometric observations.  Our statistical analysis of these 18 flares and starspots represents a fourfold increase in observations over the analysis presented in \citet{pil14}, and is presented in Table 4.  However, despite the increase in observed activity, there is little evidence for a star-planet interaction, as the measured activity does not deviate from random activity by more than 1.408$\sigma$.  

Observations that investigate the nature of the exoplanet HD 189733b and its evaporating exosphere have led to an abundance of observations near transit, at $\phi_{orbit}\sim$0.9-1.1.  Similarly, the suggestion that enhanced activity may be induced by tidal forces motivated searches at $\phi_{orbit}\sim$0.5, and the subsequent detection of some X-ray flares near this orbital phase (e.g., \citet{pil10}) spurred additional research near these orbital phases.  However, it is clear that observations recorded in the literature have poorly sampled large portions of orbital phase, particularly $\phi_{orbit}\sim$0.2-0.4.  Thus, it becomes readily apparent that the more certain orbital phases are observed, the more stellar activity is witnessed.  However, the amount of activity observed near certain orbital phases no longer seems outstanding, and the literature provides little statistical evidence for a star-planet interaction. 

\subsection{Analysis of the Entire Ensemble of Multiwavelength Observations}
Finally, the entire ensemble of existing multiwavelength observations, spanning X-ray to radio wavelengths (Tables 1 and 2) is compiled and depicted in Figure 2.  Perhaps one of the more important contributions to the present study of stellar activity on HD 189733A is that my measurement of the local minima in \emph{MOST}, APT, and Wise time series samples all orbital phases uniformly.  Thus, these measurements of the location of concentrations of active regions provide much-needed activity coverage at $\phi_{orbit}\sim$0.2-0.4, and considerably improve the statistics of stellar activity in orbital phase regions that ought to be \emph{unassociated} with star-planet interactions. 

The measured activity in each bin does not deviate from the anticipated activity by more than 0.589$\sigma$, thereby demonstrating that the observed flaring and spotting are not statistically different from an average rate across orbital phase (Table 5).  These results improve on the statistical power of the findings of \citet{pil14} in their Table 3 by a factor of 8.75.  However, these results cast doubt on the accompanying textual description presented in that paper and reproduced in Section 5.2.  Thus, this statistical analysis does not support an overdensity of stellar magnetic activity that results from either magnetic or tidal effects from the orbiting planet at $\phi_{orbit}\sim$0 \citep{cun00}, $\phi_{orbit}\sim$0.52-0.65 \citep{pil10,pil11,pil14,pil15}, $\phi_{orbit}\sim$0.8 \citep{shk08}, or as described by the spot model of \citet{mill08}.  Statistical arguments alone cast significant doubt on the existence of a star-planet interaction in the HD 189733 system.

\section{Conclusion}	
Radio observations of HD 189733A did not detect incoherent gyrosynchrotron, or coherent ECM or plasma emission, from a strong magnetic field, either in the lower corona of HD 189733A or from the auroral regions of its orbiting exoplanet, as is found at Jupiter.  Although my observations only span $\phi_{orbit}$=0.568-0.608, they do not support the hypothesis that enhanced flaring and magnetic reconnection occur at particular exoplanet orbital phases (e.g., \citet{pil10,pil11,pil14,pil15}).  Although these observations are the most sensitive to date, they can only detect powerful flares or those that emit coherent radiation.  Thus, they only weakly constrain the nature of coronal flaring on HD 189733A.

Using the published time series from \emph{MOST}, APT, and Wise Observatory, I measured the timings of photometric minima, in an effort to find active regions.  From these measurements, I computed stellar rotation and exoplanet orbital cycles using the ephemerides of \citet{far17}.  These measurements uniformly sample active region locations across orbital phases and represent a new means to investigate the effects of potential star-planet interactions.  The photometric timing analysis presented here provides much-needed data on orbital phases that have been poorly sampled previously ($\phi_{orbit}\sim$0.2-0.4).  However, the active region locations do not indicate that there is any preferred orbital phase for starspot activity.

Moreover, the multitude of observations of HD 189733A at various wavelengths provides the physical grounds to strongly constrain the properties of any star-planet interaction in this ``hot Jupiter'' system.  The many photometric starspot observations, chromospheric \ion{Ca}{2} H\&K and H$\alpha$ observations, and coronal radio and X-ray observations each provide a piece of the puzzle of the nature of activity in this system.  From these piecemeal observations, enhancement of magnetic activity at any preferred orbital phase is excluded, whether it consists of flaring, starspots, or other indicators of a persistent or comoving active region.  The coronal, chromospheric, and photometric results all exclude the theoretical notion of enhanced activity that is either magnetically ($\phi_{orbit}\sim$0.0) or tidally ($\phi_{orbit}\sim$0.0, 0.5) induced in this system as theorized by \citet{cun00}.  The enhanced flaring that \citet{shk08} alleged occurs near $\phi_{orbit}\sim$0.8 is excluded by other chromospheric observations, the coronal observations, and the photometric observations.  The comoving or persistent active region of \citet{pil10,pil11,pil14,pil15} ($\phi_{orbit}\sim$0.52-0.65) is excluded by other photospheric and chromospheric observations.  The persistent spot model presented in \citet{mill08} is excluded by other photospheric observations, by the chromospheric and coronal observations, and especially by the orbital inclination angle measurement of \citet{tri09}.  Furthermore, since the competing claims of a detected star-planet interaction in the HD 189733A/b system differ in orbital phase, they are mutually exclusive.  

This physical reasoning is buttressed by statistical analyses of the new and archival multiwavelength observations of the HD 189733 system.  My statistical analysis of the 17 photometric minima corresponding to the locations of active regions in the photosphere do not find any statistically significant enhancement in activity at any orbital phase range.  These results are echoed by an analysis of the 18 archival flares and starspots observed.  Finally, my statistical analysis of the grand ensemble of observed HD 189733A magnetic activity, spanning radio to X-ray wavelengths, demonstrates that there is no statistically significant difference betwixt the observed activity and stochastic stellar activity that is randomly distributed across the stellar surface.  Thus, statistical arguments alone indicate that stellar activity on HD 189733A is not concentrated at certain orbital phases.

Therefore, both physical and statistical arguments preclude the existence of a detectable star-planet interaction in the HD 189733A/b system; the ``on/off'' star-planet interactions in this system are decidedly \emph{off}.  Instead, HD 189733A is a very active star with heretofore sparsely sampled activity.  Furthermore, as HD 189733 is the best-studied example of the putative star-planet interaction, its absence in this system calls into question claims of star-planet interactions in other ``hot Jupiter'' systems.

The weakness or absence of the star-planet interaction in this system may help constrain the influence of HD 189733b on the system's formation and dynamical history.  Future simultaneous, multiwavelength observations of a range of magnetic activity indicators, including chromospheric H$\alpha$ and \ion{Ca}{2} H\&K, photospheric photometry, and coronal X-ray and radio emission, which densely sample stellar rotational and exoplanetary orbital phases, will better illuminate the nature of magnetism in the HD 189733 system.
 
\section{Acknowledgments}
The author would like to thank the reviewer for their careful examination of this work and their suggestions that have improved the context of its discussions.  M.R. acknowledges support from the Center for Exoplanets and Habitable Worlds and the Zaccheus Daniel Fellowship.  The Center for Exoplanets and Habitable Worlds is supported by the Pennsylvania State University and the Eberly College of Science.  At the time of the observations that are the subject of this publication, the Arecibo Observatory was operated by SRI International under a cooperative agreement with the National Science Foundation (AST-1100968) and in alliance with Ana G. M\'{e}ndez-Universidad Metropolitana and the Universities Space Research Association.

This research has made use of NASA's Astrophysics Data System.

\facility{Arecibo}
\software{IDL}

\clearpage
\begin{figure}
\centering
\includegraphics[width=0.8\textwidth,angle=0]{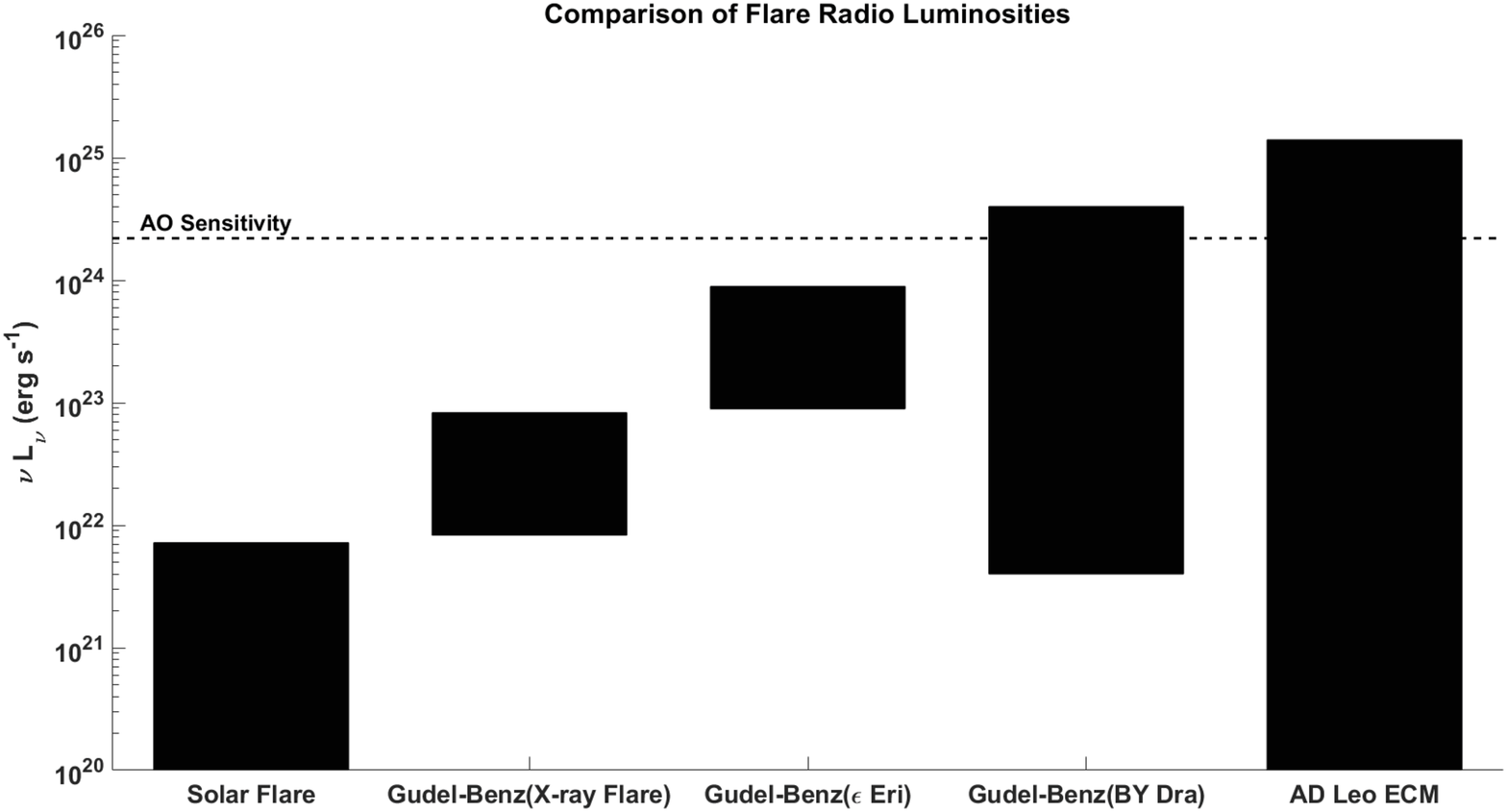}
\caption{Comparison of $\nu L_{\nu}$ radio luminosities for various stellar flare scenarios.  ``G\"{u}del-Benz (X-Ray Flare)'' denotes the radio luminosity range that corresponds to the peak X-ray flare luminosity observed by \citet{pil14}, as computed from the empirical G\"{u}del-Benz relationship.  ``G\"{u}del-Benz ($\epsilon$ Eri)'' denotes the radio luminosity range that corresponds to \emph{Chandra}/LETGS X-ray emission from HD 189733A spectral type analog $\epsilon$ Eri \citep{nes02}.  Similarly, other BY Dra-classified stars also emit a range of radio luminosities \citep{ben10}.  The radio observations described in Section 3.1 are sensitive to large, incoherent, gyrosynchrotron flares and modest coherent, beamed, ECM flares similar to those that have been observed on AD Leo. The dashed line demarcates the sensitivity of Arecibo Observatory to radio emission from flares on HD 189733A.} 
\end{figure}

\clearpage
\begin{deluxetable}{lcccccc}
\tabletypesize{\scriptsize}
\tablecolumns{7}
\tablewidth{0pt}
\tablecaption{Photometric Activity Measurements and Phases}
\tablehead{
	\colhead{Observation}&
	\colhead{Observation}&
	\colhead{Active Region}&
	\colhead{Active Region}&
	\colhead{Active Region}&
	\colhead{Source}&
	\colhead{Time Series}\\
	\colhead{Start (JD)}&
	\colhead{Finish (JD)}&
	\colhead{Midpoint (JD)}&
	\colhead{Rotation Cycle}&
	\colhead{Orbit Cycle}&
	\colhead{}&
	\colhead{Reference}	
}
\startdata
2453949.00 & 2453962.00 & 2453954.95 & 27.13 & 146.74 & \emph{MOST} & 1\\
2453962.00 & 2453969.00 & 2453966.90 & 28.16 & 152.29 & \emph{MOST} & 1\\ 
2454299.00 & 2454311.00 & 2454302.91 & 56.13 & 303.58 & \emph{MOST} & 2\\
2454311.00 & 2454321.00 & 2454313.94 & 57.05 & 308.55 & \emph{MOST} & 2\\
2454321.00 & 2454325.00 & 2454322.16 & 57.73 & 312.26 & \emph{MOST} & 2\\
2454325.00 & 2454329.80 & 2454325.63 & 58.02 & 313.82 & \emph{MOST} & 2\\
2455130.00 & 2455136.00 & 2455131.88 & 125.21 & 677.23 & APT\&Wise & 3\\
2455136.00 & 2455141.00 & 2455138.00 & 125.72 & 679.99 & APT\&Wise & 3\\
2455141.00 & 2455148.00 & 2455142.19 & 126.07 & 681.88 & APT\&Wise & 3\\
2455148.00 & 2455152.00 & 2455149.69 & 126.69 & 685.26 & APT\&Wise & 3\\
2455152.00 & 2455158.00 & 2455154.50 & 127.09 & 687.43 & APT\&Wise & 3\\
2455158.00 & 2455164.00 & 2455160.94 & 127.63 & 690.33 & APT\&Wise & 3\\
2455164.00 & 2455171.25 & 2455165.63 & 128.02 & 692.44 & APT\&Wise & 3\\
2455321.43 & 2455328.00 & 2455327.00 & 141.47 & 765.18 & APT\&Wise & 3\\
2455328.00 & 2455334.00 & 2455333.44 & 142.00 & 768.08 & APT\&Wise & 3\\
2455334.00 & 2455341.00 & 2455340.00 & 142.55 & 771.04 & APT\&Wise & 3\\
2455341.00 & 2455347.00 & 2455345.25 & 142.99 & 773.41 & APT\&Wise & 3\\
\enddata
\tablecomments{These temporal measurements represent local minima in the time series of the photometric flux from HD 189733A. See Section 3 for more details on the measurement of parameters and the computation of rotation and orbital cycles. Starspots are correlated on rotational timescales, but not exoplanet orbital timescales. {\bf Time series references:} (1) \citet{mill08}; (2) \citet{boi09}; (3) \citet{sing11}.  Reference 2 time series are in BJD.}
\end{deluxetable}

\clearpage
\begin{deluxetable}{llllll}
\tabletypesize{\scriptsize}
\tablecolumns{6}
\tablewidth{0pt}
\tablecaption{Chronological Listing of HD 189733A Activity Reported in the Literature}
\tablehead{
	\colhead{Date}&
	\colhead{Wavelength}&
	\colhead{Activity Indicator}&
	\colhead{Orbital Phases}&
	\colhead{Activity Orbital} &
	\colhead{References}\\
	\colhead{}&
	\colhead{Region}&
	\colhead{}&
	\colhead{Monitored}&
	\colhead{Phase}&
	\colhead{}
}
\startdata
2006 May 22 & Optical & Broadband (5500-10500\AA) & 0.932-1.101 & 0.002 & \citet{pon07}\\
2006 May 26 & Optical & Broadband (5500-10500\AA) & 0.915-1.047 & 0.007 & \citet{pon07}\\
2006 Jun & Optical & Ca II H\&K (3700-10400\AA) & 0.20-0.80\tablenotemark{a} & 0.80(X) & \citet{shk08}\\
2006 Jul 14\tablenotemark{b} & Optical & Broadband (5500-10500\AA) & 0.926-1.058 & 0.010 & \citet{pon07}\\
2007 Apr 17 & X-Ray & 0.3-8 keV & 0.84-1.13 & 0.016(X),0.87?\tablenotemark{c} & \citet{pil10}\\
2007 Apr 21 & Radio & 307-347 MHz & 0.444-0.557 & N/A & \citet{smi09}\\
2007 Jun 10 & UV & CII (1150-1750\AA) & 0.958-1.061 & N/A & \citet{lec10}\\
2007 Jun 18 & UV & CII (1150-1750\AA) & 0.952-1.055 & N/A & \citet{lec10}\\
2008 Apr 24 & UV & CII (1150-1750\AA) & 0.967-1.039 & 0.0 & \citet{lec10}\\
2008 Aug 14 & Radio & 241-247; 598-630 MHz & 0.448-0.606 & N/A & \citet{lec09}\\
2009 May 18-19 & X-Ray & 0.3-8 keV & 0.45-0.62 & 0.45?,0.54 & \citet{pil10}\\
2009 Aug 15 & Radio & 140-156 MHz & 0.410-0.581 & N/A & \citet{lec11}\\
2009 Nov 20 & UV/Optical & Broadband (2900-5700\AA) & 0.965-1.040 & 0.992 & \citet{sing11}\\
2010 May 18 & UV/Optical & Broadband (2900-5700\AA) & 0.956-1.032 & 0.995? & \citet{sing11}\\
2011 Apr 30 & X-Ray & 0.3-1.5 keV & 0.405-0.608 & 0.52 & \citet{pil11}\\
2011 Jul 5 & X-Ray & 0.1-2 keV & 0.943-1.043 & N/A & \citet{pop13}\\
2011 Jul 12 & X-Ray & 0.1-2 keV & 0.938-1.042 & N/A & \citet{pop13}\\
2011 Jul 16 & X-Ray & 0.1-2 keV & 0.935-1.029 & N/A & \citet{pop13}\\
2011 Jul 18 & X-Ray & 0.1-2 keV & 0.938-1.041 & N/A & \citet{pop13}\\
2011 Jul 21 & X-Ray & 0.1-2 keV & 0.938-1.041 & N/A & \citet{pop13}\\
2011 Jul 23 & X-Ray & 0.1-2 keV & 0.951-1.054 & N/A & \citet{pop13}\\
2011 Sep 7 & Radio & 4.25-5.25 GHz & 0.568-0.608 & N/A & This work\\
2011 Sep 7 & X-Ray & 0.3-3 keV & 0.70-1.23 & 0.84 & \citet{lec12}\\
2012 May 7 & X-Ray & 0.3-2.5 keV & 0.45-0.75 & 0.57,0.64 & \citet{pil14}\\
2012 Jul 4 & Optical & Ca II H\&K (3891-4011\AA) & 0.96-1.05 & 0.002 & \citet{cze15}\\
2013 Jul 4 & Optical & H$\alpha$ (6563\AA) & 0.956-1.038 & Absorption? & \citet{cau17a}\\
2013 Sep 12 & UV & CII, SiII/III/IV (1150-1450\AA) & 0.500-0.626 & 0.525,0.588 & \citet{pil15}\\
2015 Aug 4 & Optical & H$\alpha$ (6563\AA) & 0.937-1.063 & Absorption? & \citet{cau17a}\\
2016 Jul 29 & Optical & H$\alpha$ (6563\AA) & 0.100-0.250 & N/A & \citet{cau17a}\\
2016 Jul 30 & Optical & H$\alpha$ (6563\AA) & 0.570-0.608 & N/A & \citet{cau17a}\\
2016 Jul 31 & Optical & H$\alpha$ (6563\AA) & 0.019-0.077 & N/A & \citet{cau17a}\\
2016 Aug 01 & Optical & H$\alpha$ (6563\AA) & 0.450-0.556 & N/A & \citet{cau17a}\\
2016 Sep 19 & Optical & H$\alpha$ (6563\AA) & 0.510-0.613 & N/A & \citet{cau17a}\\
\enddata
\tablecomments{Multiwavelength stellar activity, including flares and starspots, reported in the literature, but excluding the photometrically determined active regions listed in Table 1.  Nondetections are listed when they are discussed in the text.  Activity indicator properties are listed in the energy (X-rays), frequency (radio), or wavelength (UV, optical) formats appropriate to their domains.  Question marks in the ``Activity Orbital Phase'' column denote potential flare events that were not described in the reference provided but are indicated by the supporting data.  Observed H$\alpha$ variability listed as ``Absorption?'' likely is the result of circumstellar material from the evaporating exosphere but may denote stellar activity.  All listed broadband observations are starspot detections.  The phase values for \citet{lec10}, \citet{pil11}, \citet{pop13}, and the presented radio observations represent exact MJDs converted to HJDs.  The remaining observations are estimates.  MJD-to-HJD corrections introduce a shift in $\phi_{orbit}$ of $\pm$0.002.}
\tablenotetext{a}{\citet{shk08} observations are modeled as four 2 hr epochs centered on the orbital phases reported in their Figure 12 ($\phi_{orbit}$=0.19-0.21, 0.29-0.31, 0.64-0.66, and 0.79-0.81). The claim of enhanced activity at $\phi_{orbit}$=0.80 actually represents a large decline in chromospheric emission; thus, this event is marked (X) and should be discounted in future investigations of activity from this star.}
\tablenotetext{b}{The observations described in Section 2 of \cite{pon07} should start at JD 2453877.718, not JD 245877.718 as printed.}
\tablenotetext{c}{\citet{pil10} report a flare near ($\phi\sim$0) based on a hardening spectrum.  However, the count rates associated with this event in their Figure 2 do not make this flare distinguishable from other background activity; hence, its detection is marked (X) and it is not included in the statistical analysis.  Alternatively, an obvious flare exists at $\phi_{orbit}\sim$0.87 that is not mentioned.}
\end{deluxetable}

\clearpage
\begin{deluxetable}{lcccc}
\tabletypesize{\scriptsize}
\tablecolumns{5}
\tablewidth{0pt}
\tablecaption{Statistical Analysis of \emph{MOST}, APT, and Wise Photometry}
\tablehead{
	\colhead{Orbital Phase}&
	\colhead{Cumulative Observing}&
	\colhead{Expected}&
	\colhead{Measured} &
	\colhead{Significance}\\
	\colhead{Range}&
	\colhead{Fraction}&
	\colhead{Activity}&
	\colhead{Activity}&
	\colhead{($\sigma$)}
}
\startdata
0.901-0.100 & 0.2 & 3.4 & 3.0 & 0.22\\
0.101-0.300 & 0.2 & 3.4 & 5.0 & 0.87\\
0.301-0.500 & 0.2 & 3.4 & 4.0 & 0.33\\
0.501-0.700 & 0.2 & 3.4 & 2.0 & 0.76\\ 
0.701-0.900 & 0.2 & 3.4 & 3.0 & 0.22\\
\enddata
\tablecomments{Statistical analysis of active region locations determined by photometric minima in Table 1.  The computed results do not represent a statistically significant deviation from active regions that are randomly distributed in orbital phase.  Star-planet interactions appear to have little or no influence on the photospheric activity of HD 189733A.}
\end{deluxetable}

\begin{deluxetable}{lcccc}
\tabletypesize{\scriptsize}
\tablecolumns{5}
\tablewidth{0pt}
\tablecaption{Statistical Analysis of HD 189733A Activity in the Literature}
\tablehead{
	\colhead{Orbital Phase}&
	\colhead{Cumulative Observing}&
	\colhead{Expected}&
	\colhead{Measured} &
	\colhead{Significance}\\
	\colhead{Range}&
	\colhead{Fraction}&
	\colhead{Activity}&
	\colhead{Activity}&
	\colhead{($\sigma$)}
}
\startdata
0.901-0.100 & 0.510 & 9.18 & 8.0 & 0.389\\
0.101-0.300 & 0.078 & 1.40 & 0.0 & 1.185\\
0.301-0.500 & 0.105 & 1.89 & 1.0 & 0.646\\
0.501-0.700 & 0.232 & 4.17 & 6.0 & 0.896\\ 
0.701-0.900 & 0.075 & 1.36 & 3.0 & 1.408\\
\enddata
\tablecomments{Statistical analysis of the flares and starspots reported in the literature (Table 2), excluding the new radio and photometric data presented in this work.  While this analysis finds somehwat greater statistical support for enhanced activity at $\phi_{orbit}$=0.5-0.7 and $\phi_{orbit}$=0.7-0.9 bins than did \citet{pil14}, they are still not statistically significant.}
\end{deluxetable}
\clearpage

\begin{deluxetable}{lcccc}
\tabletypesize{\scriptsize}
\tablecolumns{5}
\tablewidth{0pt}
\tablecaption{Statistical Analysis of All HD 189733A Activity}
\tablehead{
	\colhead{Orbital Phase}&
	\colhead{Cumulative Observing}&
	\colhead{Expected}&
	\colhead{Measured} &
	\colhead{Significance}\\
	\colhead{Range}&
	\colhead{Fraction}&
	\colhead{Activity}&
	\colhead{Activity}&
	\colhead{($\sigma$)}
}
\startdata
0.901-0.100 & 0.263 & 9.21 & 11.0 & 0.589\\
0.101-0.300 & 0.175 & 6.11 & 5.0 & 0.449\\
0.301-0.500 & 0.180 & 6.30 & 5.0 & 0.519\\
0.501-0.700 & 0.208 & 7.28 & 8.0 & 0.266\\ 
0.701-0.900 & 0.174 & 6.09 & 6.0 & 0.038\\
\enddata
\tablecomments{Statistical analysis of all flare and spot observations reported in the literature and this work, that result from activity across X-ray, UV, optical, and radio wavelengths (c.f., \citet{pil14}).  The orbital phase ranges were uniformly divided in phase, starting with 0.901-0.100, which corrsponds to placing all transit observations into the same bin.  Expected flares were computed assuming uniform activity across orbital phase.  No statistically significant overdensities of activity were found among the orbital phase ranges surveyed.  If the two H$\alpha$ observations tentatively marked as ``Absorption?'' caused by obscuring material from HD 189733b's exosphere are actually manifestations of chromospheric activity, the listed stellar activity will even more closely match that anticipated for uniform, stochastic flaring across the stellar disk.}
\end{deluxetable}

\begin{figure}
\centering
\includegraphics[width=0.8\textwidth,angle=0]{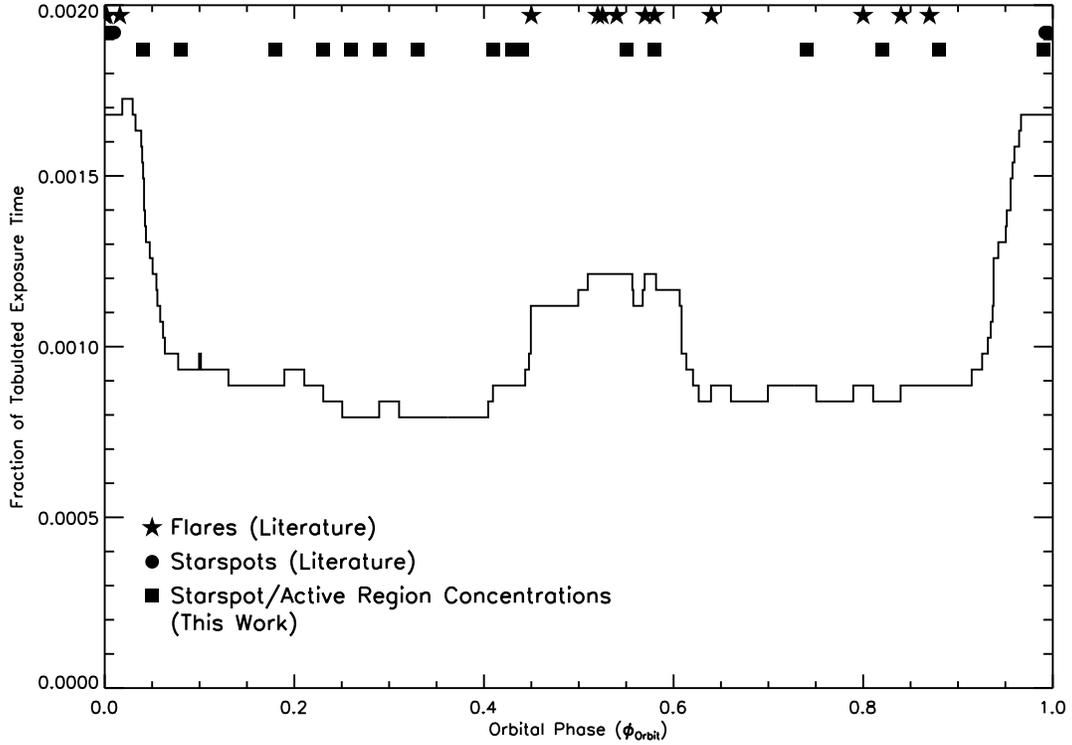}
\caption{Stellar activity as a function of orbital phase.  The y-axis lists the cumulative fractional exposure time of multiwavelength observations for bins of 0.001 temporal width in orbital phase.  The orbital phases of observed stellar flares, starspots, and significant active regions are marked at the top of the diagram by filled stars, circles, and squares, respectively.  Note that many data were collected during transit ($\phi_{orbit}\sim$0.9-1.1), resulting in three overlapping flares and five overlapping starspot detections.  The photometric determination of active regions presented in this work (Table 1) significantly improves the coverage of activity at $\phi_{orbit}\sim$0.2-0.4.  See Tables 1 and 2 for specific activity properties.}
\end{figure}

\end{document}